# Characterization of Using Hybrid Beamforming in mmWave Virtual Reality

Nasim Alikhani, Abbas Mohammadi,
AUT-Wireless Research Lab.
Electrical Engineering Department,
Amirkabir University of Technology
Tehran, Iran

*abm125@aut.ac.ir*

*Abstract*—Wireless Virtual Reality (VR) is increasingly in demand in Wireless LANs (WLANs). In this paper, a utility function for resource management in wireless VR is proposed. Maximizing the sum rate metric in transmitting VR audio or videos is an important factor for ascertaining low latency in obtaining QoS requirement of users in VR, so forth mmWave frequency band in WLAN technology should be used. This frequency band is presented in IEEE 802.11ad/ay. Resource access method in IEEE 802.11ay standard is Multi-User MIMO (MU-MIMO) with OFDM modulation. Operating at mmWave frequency band is equal to use massive number of antenna to enhance the received power in (Line of Sight) LoS direction by inducing sever propagation with small wavelength. Also for reducing the complexity of hardware in mmWave technology, designers should select some number of connected phase shifters to each antenna element by hybrid beamforming method. Processing delay, transmission delay and queue delay should be considered in acquiring QoS metric in terms of utility function. The optimal closed form expression of the multi-attribute utility function is based on these delays that are calculated by downlink and uplink rates in assistant with hybrid beamforming. Trends of transmission delay and multi-attribute utility function in various $E_s/N_0$ values and different scenarios are analyzed. Based on these results, 95.4% accuracy in comparison with ns3 in uplink and downlink channel modeling in IEEE 802.11ay standard's indoor environment has been reported. Also, it is shown that min channel gain consideration can cause reduction in the value of the utility function and incursion in transmission delay in VR.

*Index Terms*— Virtual Reality (VR), QoS, mmWave, IEEE802.11ay, MU-MIMO-OFDM, utility function, hybrid beamforming.

## 1.1. INTRODUCTION

Virtual reality (VR) technology provides a real-time immersive experience that tightly blends physical and digital reality. This technology is obtaining attention due to its ability to present an immersive viewing experience to users [1-6]. VR services are designed to build a virtual environment to mimic the real world and immerse participants in virtual worlds. Although the quality of the experience can be improved by changing the frame size and frame rate, there are limitations on the values of these parameters in the wireless environment [7,8]. On the other hand one of the main problems in wired VR application is the restriction of mobility of VR devices. The second problem is that VR application is delay-sensitive. To overcome these limitations, VR services can be supported with wireless cellular connectivity that may provide better user experiences [9,10].

A specific feature of heavy traffic load on VR applications in wireless network is delay sensitivity. The current Wi-Fi standards in Wireless Local Area Networks (WLAN) are unable to guarantee the desired Quality of Service (QoS) requirements, especially in dense environments [11]. On the other hand, mmWave technology can provide a high transmission rate, low latency, and high transmission reliability. Built upon 802.11ad, IEEE 802.11ay aims to offer about 100 Gbps throughput, ultra-low latency by using technological advancements such as MIMO communication, channel bonding/aggregation, and new beamforming techniques. Various VR services have different traffic loads, delay, and bandwidth requirements [12]. IEEE 802.11ay standard [13] works on the millimeter-wave frequency range and uses MU-MIMO access method in Downlink (DL) with OFDM modulation. OFDM modulation in IEEE 802.11ay increases the rate in DL by using a various number of subcarriers [14-16]. In MU-MIMO access mechanism, some users receive services simultaneously from a transmitter such as Access Point (AP). AP has to schedule users in a parallel manner to maximize the overall throughput. In order to realize the benefit of the MU-MIMO access mechanism and guarantee the required QoS, it is essential to acquire updated Channel State Information (CSI) from all the users. Hence, there are a trade-off between the efficiency of the scheduler and the CSI overhead. Generally, AP limits the number of users based on CSI feedback. The best user CSI and suitable channels need to be obtained before the user is scheduled [17].

Meanwhile, in order to overcome the severe propagation loss of mmWave signals, large antenna arrays in configuration of mmWave systems should be used [18-21]. The usage of a large number of antennas in mmWave systems is equal to a large amount of RF chain that has great power consumption which makes full digital beamforming impractical. Because of this limitation, hybrid beamforming can be used in mmWave systems. Hybrid beamforming method divides the beamforming method into digital and analog domains [10, 23]. So in order to increase the energy efficiency in MU-MIMO mmWave systems and reduce the number of RF chains at the cost of only slight performance degradation, the combination of a low-dimensional digital baseband beamformer and a high-dimensional analog beamformer is essential [18].

In addition, important requirements on VR service are tracking accuracy, transmission delay, processing delay and queue delay [24]. Authors in [25-28] have considered utility function with less features that are almost dependent to each other. Tracking accuracy, transmission delay, processing delay and queue delay affect user experience. They are not independent from each other on affecting requirements of QoS for users. In order to consider them jointly, we propose a multi-attribute utility function that consider all of them based on the computed rates in Downlink (DL) and Uplink (UL) with hybrid beamforming and infinite buffer size in the transmitter in DL. For the hybrid beamforming method in this paper, only once estimation of matrices in subcarrier frequencies (without iteratively estimating beamforming matrices) is needed. Then, we explore hybrid beamforming design for MU-MIMO OFDM systems over mmWave channels. Finally, by computing the transmission delay, processing delay and queue delay at users and AP side, the value of the utility function is evaluated. To the best of our knowledge, this is the first method in VR applications in IEEE 802.11ay that uses the multi-attribute utility function with considering transmission delay, processing delay, and queue delay. The main contributions of our paper are summarized as follows,

1) Proposed hybrid beamforming method has low complexity in calculation of beamforming matrices, because they are computed without iterative methods.

2) A multi attribute utility function by considering delay, transmission and processing delay is proposed.

3) The results with ns3 (network simulator) with consideration of IEEE 802.11ay specifications in the PHY layer are extracted. Moreover, the results of MATLAB simulation are compared with ns3 results.

4) A lower bound for DL channel is calculated based on subcarriers and the value of utility function is calculated and compared with the utility function for DL channel by considering the mean gain values of channel in all subcarriers. So we have considered two scenarios as mean channel gain and min channel gain in the case of subcarriers and compare their results of utility function and transmission delay line.

5) The results of the proposed optimal closed form expression of multi-attribute utility function in MATLAB and ns3 simulations are compared with six different codebooks that are generated in ns3: "two-Antenna, one-RF chain", "two-Antenna, two-RF chain", "four-Antenna, one-RF chain", "four-Antenna, two-RF chain", "eight-Antenna, one-RF chain", "eight-Antenna, two-RF chain".

The rest of the paper is organized as follows. In Section 1.2, system model for channels in the uplink and downlink are presented. In section 1.3, delay considerations are explained. Also the routine for discovering the channel parameters of downlink is described. In section 1.4 the optimization problem is analyzed. In section 1.5, the indoor area and simulation results are explained. In the last section, conclusion is expressed.

## 1.2. SYSTEM MODEL

In this section formulas for UL and DL channels, delay purposes and construction of utility function are explained. The system model is shown in Fig. 1, where DL and UL channels between each user and AP and intra-interference in DL and UL are shown in Fig. 1. In this Fig, the black solid lines represent the main lobe beam direction in UL from user to AP in *n*th subcarrier index. These channels are $h_{1,1,n}^{UL}, h_{2,2,n}^{UL}$ in *n*th subcarrier. The black continues dotted lines represent the main lobe beam direction from each antenna in DL from AP to each user. These channels are shown by $h_{1,1}^{DL}, h_{2,2}^{DL}$. The grey dotted lines represent the inter interference in DL direction from AP to user. These interference channels are $h_{2,1}^{DL}, h_{1,2}^{DL}$. The grey solid lines represent the inter interference in UL direction from user to AP. These interference channels are denoted by $h_{2,1,n}^{UL}, h_{1,2,n}^{UL}$.

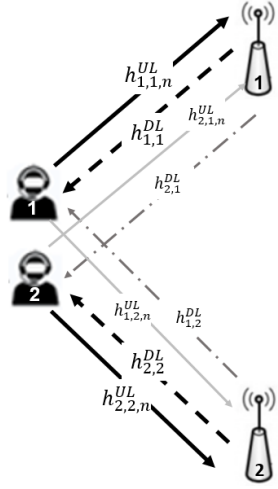

**Fig. 1.** System Model configuration.

*1.2.1. UL channel*

The frequency band that is considered in UL channel can be sub-6GHz. The channel matrix in the UL is as below:

$$h_{ijn}^{UL}(x_i, y_i, z_i) = g\left(d_{ij}(x_i, y_i, z_i)\right)^{-w} \quad (1)$$

Where channel matrix is denoted as $h_{ijn}^{UL}$ for $i$th user, $j$th AP in $n$th subcarrier. $g$ is the complex Gaussian channel gain in each subcarrier. $d_{ij}$ is the Euclidean vector distance between $i$th user and $j$th AP (each user and AP has 3D vector for position). $w$ is the pathloss exponent value. The 3D coordinate position of $i$th user is $(x_i, y_i, z_i)$.

The complex Gaussian channel matrix $g$ in uplink is considered as a complex Gaussian matrix with unit amplitude and different phase values that are related to subcarrier frequencies as given by equation (2),

$$g = \exp\left(j\left(\frac{\pi}{180} : \frac{\pi}{180} : \frac{\pi}{180} \times N_{SC}\right)\right) \quad (2)$$

Where $N_{SC}$ is the number of subcarriers. So UL channel is $h_{ij}^{UL} = \sum_{N_{SC}} h_{ijn}^{UL}$.

*1.2.2. DL channel*

The frequency band that is considered in DL channel is mmWave technology. $N_t$ is the number of transmit antenna at each AP. $N_r$ is the number of receive antenna for each user. $N_{DS}$ is the number of required bit stream for each user. $N_{RF}$ is the number of RF chain in antenna configuration of AP.

1- *Calculation of the parameters in DL channel*

Delay is assumed as $\tau = \frac{d}{c}$ in DL where $d$ is the distance between each AP and user. $c$ is the speed of light. Path gain at each subcarrier is denoted as "$pg_{ij}$" (between $i$th user and $j$th AP). $\lambda$ is the wave length by considering that carrier frequency is equal to 60 GHz. $20 \log_{10}\left(\frac{\lambda}{4\pi d_{ij}}\right)$ is the free space path-loss value in dB.

$$pg_{ij} = 20 \log_{10}\left(\frac{\lambda}{4\pi d_{ij}}\right) \times \exp\left(j\left(\frac{\pi}{180} : \frac{\pi}{180} : \frac{\pi}{180} \times N_{SC}\right)\right) \quad (3)$$

By considering the position of AP as TX and position of user as RX notations in DL, the values of DOD (Direction of Departure) =RX-TX and DOA (Direction of Arrival) =-DOD can be computed. By using the values of DOA and DOD, AOA (Angle of Arrival) in azimuth direction and AOD (Angle of Departure) in azimuth direction are computed.

$$AOD_{AZ} = mod\left(tg^{-1}\left(\frac{DOD(2)}{DOD(1)}\right), 360\right) \quad (4)$$

$$AOA_{AZ} = mod\left(tg^{-1}\left(\frac{DOA(2)}{DOA(1)}\right), 360\right) \quad (5)$$

DOA(2) and DOA(1) are the second and first dimension of DOA respectively. Similarly, DOD(2) and DOD(1) are the second and first dimension of DOD respectively. As a consequence $DOA_{AZ}$, $AOA_{AZ}$ are converted from radians to degrees and stored respectively as $\theta_{aod_{AZ}}, \phi_{aoa_{AZ}}$. ULA (Uniform Linear Array) antenna configuration is considered on both the user and AP sides. The steering vector for this antenna type is shown below:

$$AOD_{AZ} = \frac{1}{\sqrt{N_t}} \exp\left(-\frac{j[0:N_t-1]2\pi d}{\lambda} \sin(\theta_{aod_{AZ}})\right) \tag{6}$$

$$AOA_{AZ} = \frac{1}{\sqrt{N_r}} \exp\left(-\frac{j[0:N_r-1]2\pi d}{\lambda} \sin(\phi_{aoa_{AZ}})\right) \tag{7}$$

So complex channel in DL can be evaluated as equation (8):

$$h_{ij}^{DL} = \sum_t \sum_{N_{SC}} \left(10^{\frac{pg_{ij}}{10}}\right) \times AOD_{AZ} \times AOA_{AZ} \times \exp(-t/\tau) \tag{8}$$

The variable $t$ in equation (8) is the time counter in processing the DL channel $h_{ij}^{DL}$. This channel is equal to the summation of complex values generated from components of delay between each user and AP, AOA from each user and AOD from each AP and path-gain over various times and subcarriers.

2- *Beamforming*

The beamforming method is used to lower the hardware cost in mmWave systems. In this section, the mathematical equations for both full digital beamforming and hybrid beamforming are provided. We represent that the channel matrix in each subcarrier frequency is denoted as $H_{SC} \in \mathbb{C}^{N_r \times N_{DS}}$.

a- *Full digital beamforming*

In full digital beamforming, each RF chains is connected to all antenna elements. In the first step, the channel $H_{SC}$ is decomposed by SVD (Singular Value Decomposition) as: $[U\ W\ V] = svd(H_{SC})$. The digital precoder in each subcarrier frequency can be estimated as follows:

$$P_{D_{SC}} = [U \times [I_{N_{DS}}\ \ 0_{N_{DS} \times (N_t - N_{DS})}]] \tag{9}$$

Where $I_{N_{DS}}$ is an identity matrix. $0_{N_{DS} \times (N_t - N_{DS})}$ is a zero matrix that is used for dimensions with a fixed size in a specific number of bit streams and the number of antennas at each AP. U and V are unitary matrices. So $P_{D_{SC}} \in \mathbb{C}^{N_t \times N_{DSS}}$ is a complex semi-unitary matrix.

The digital combiner matrix is denoted by $G_{D_{SC}} \in \mathbb{C}^{N_r \times N_{DS}}$ and is defined by equation (10):

$$G_{D_{SC}} = [V \times [I_{NDS}\ \ 0_{N_{DS} \times (N_r - N_{DS})}]] \tag{10}$$

Analog beamforming matrices are considered as a constant value in each subcarrier frequency.

b- *Estimation of analog beamforming matrices*

Initially, the sum of the values of channel matrix in all subcarrier frequencies is computed, followed by the computation of the SVD from this matrix. To calculate the analog combiner matrix, $H_{SC}H_{SC}^H$ should be computed first which has dimensions of $N_r \times N_r$ (because the complex channel $H_{SC}$ has dimensions of $N_r \times N_t$), and then the sum of values of the matrix $H_{SC}H_{SC}^H$ in terms of all subcarrier frequencies is calculated.

$$H_{S_1} = \sum_{SC=1}^{N_{SC}} H_{SC} H_{SC}^H \tag{11}$$

Where $N_{SC}$ is the total number of subcarriers.

Then the decomposition of $H_{S_1} = U_{S_1} W_{S_1} V_{S_1}^H$ is calculated by SVD. The analog combiner matrix is $G_A = \frac{U_{S_1}}{|U_{S_1}|}$, where $G_A \in \mathbb{C}^{1 \times N_r}$.

To estimate the analog precoder, the matrix $H_{SC}^H H_{SC}$ which has dimensions equal to $N_t \times N_t$, is calculated in each subcarrier, and then it is denoted as $H_{S_2}$.

$$H_{S_2} = \sum_{SC=1}^{N_{SC}} H_{SC}^H H_{SC} \tag{12}$$

Then the decomposition of $H_{S_2} = U_{S_2} W_{S_2} V_{S_2}^H$ by SVD is calculated. The analog precoder matrix is represented by $P_A$ which is equal to $P_A = \frac{U_{S_2}}{|U_{S_2}|}$, where $P_A \in \mathbb{C}^{N_t \times N_{RF}}$.

c- *Hybrid beamforming for calculating digital beamforming matrix*

The difference between full digital beamforming and hybrid beamforming is that in the hybrid method, each antenna is connected to a series of phase shifters due to the high energy consumption in the antenna configuration in the mmWave frequency band. Therefore, the channel matrix should be updated by considering the analog beamforming matrix and then using their values for calculating the digital beamforming [16].

As a consequence, $H_{D_{SC}} = G_A^H H_{SC} P_A$ can be concluded. The channel matrix in baseband in each subcarrier is denoted as $H_{eff_{SC}}$ and is equal to $H_{eff_{SC}} = G_{D_{SC}}^H H_{D_{SC}} P_{D_{SC}}$.

So, for computing $G_{D_{SC}}, P_{D_{SC}}$ in each subcarrier, the SVD of channel $H_{D_{SC}}$ is computed as $H_{D_{SC}} = U_{D_{SC}} W_{D_{SC}} V_{D_{SC}}^H$.
The digital beamforming matrices (digital precoder and digital combiner) in each subcarrier frequency are respectively shown as below:

$$\hat{P}_{D_{SC}} = V_{D_{SC}} \times [I_{N_{DS}} \; 0_{N_{DS} \times (N_{RF} - N_{DS})}] \tag{13}$$

$$\hat{G}_{D_{SC}} = U_{D_{SC}} \times [I_{N_{DS}} \; 0_{N_{DS} \times (1 - N_{DS})}] \tag{14}$$

Due to equation (14), the number of RF chains on the user side is 1. Consequently, these equalities can be concluded: $\hat{G}_{D_{SC}} = \hat{G}_{D_{SC(i,n)}}, \hat{P}_{D_{SC}} = \hat{P}_{D_{SC(i,n)}}$. The digital precoder depends on power of the transmitter, therefore, normalizing the digital precoder by the power of the AP in the DL is satisfactory.

In this hybrid beamforming method, by once estimating matrices $H_{S_1}, H_{S_2}, H_{D_{SC}}$ across all subcarrier frequencies (without iteratively estimating beamforming matrices), beamforming matrices are discovered by the SVD of channel matrices.

To discover the DL rate for each user in each subcarrier, it should be noted that inter and intra interferences has been computed. Intra interference occurs within the user's own cell, while Inter interference occurs between cells.

SINR in UL is equivalent to:

$$SINR_{ijn}^{UL} = \frac{P_{U_i}(h_{ijn}^{UL})^2}{\sigma^2 + \sum_{l \neq i} P_{U_l}(h_{ljn}^{UL})^2 + \sum_{b=1, b \neq j}^{B} \sum_{k=1, k \neq i}^{U_b} P_{U_k}(h_{kbn}^{UL})^2} \tag{15}$$

SINR in DL is equivalent to:

$$SINR_{ij}^{DL} = \frac{P_{B_j}(h_{ij}^{DL})^2}{\sigma^2 + \sum_{l \neq i} P_{B_j}(h_{lj}^{DL})^2 + \sum_{b=1, b \neq j}^{B} \sum_{k=1, k \neq i}^{U_b} P_{B_b}(h_{kb}^{DL})^2} \tag{16}$$

Where $P_{U_i}$ is the power of the $i$th user. $U_b$ is equal to users in the coverage of $b$th AP. $B$ is the number of APs. $P_{B_j}$ is the power of the $j$th AP. $\sigma^2$ is the power of complex Gaussian noise, which is independent of inter and intra interference signals.

The UL rate and DL rate are respectively equivalent to:

$$c_{ijn}^{UL} = BW \times \log_2(1 + SINR_{ijn}^{UL}) \tag{17}$$

$$c_{ij}^{DL} = BW \times \log_2(1 + SINR_{ij}^{DL}) \tag{18}$$

### 1.3. DELAY COMPUTATION AND UTILITY FUNCTION

This section discusses the computations of delay and the utility function.

*1.3.1 Delay*

In considering delays, three factors have been presented: 1-delay in UL, 2-delay in DL and 3-queue delay. The transmission delay for the $i$th user, $j$th AP in the $n$th subcarrier frequency is given by equation (19),

$$D_{ijn}^T = \frac{S_i}{c_{ij}^{DL}} + \frac{A_i}{c_{ijn}^{UL}} \tag{19}$$

$S_i$ represents the maximum number of bits in DL that each AP transmit to the $i$th user. While $A_i$ denotes the size of the tracking vector in UL that each user transmits to the specified AP.
The delay for processing information is as follows:

$$D_i^p(K_{ijn}) = \frac{v\left(l\left(X_i(SINR_{ijn}^{UL})\right), l(X_i^R)\right)}{\frac{M}{N_j}} \tag{20}$$

Here, $0 \leq v\left(l\left(X_i(SINR_{ijn}^{UL})\right), l(X_i^R)\right) \leq S_i$, where $v$ is the number of bits in transmitting the image from $l\left(X_i(SINR_{ijn}^{UL})\right)$ to $l(X_i^R)$, calculated by each AP. $M$ represents the overall processing limit of each AP. $N_j$ is the dedicated power to each user from the $j$th AP, and $K_{ijn}$ signifies the accuracy in routing the $i$th user in the $n$th subcarrier by the $j$th AP.

Regarding the queue delay issue, the distribution of users requests follows a poisson distribution with a mean $\lambda_i$. The servicing time for the request of each user has an exponential distribution with parameter $\mu_j$. The buffer size in each AP is considered infinite.

The inequality $\mu_j > \lambda_i$ is assumed for the tradeoff between the time of user's requests and servicing time. The delay time for the queue is represented as $\frac{1}{\mu_j - \lambda_i}$ and total delay time is given by equation (21),

$$D_{ijn} = D_i^p(K_{ijn}) + D_{ijn}^T + \left(\frac{1}{\mu_j - \lambda_i}\right) \tag{21}$$

### 1.3.2. Utility Function

The multi attribute utility function in this paper has a joint function takes a conditional form with respect to the accuracy of routing [24]. This function describes the utility function between the $i$th user, $j$th AP as $U_i(D_{ijn}, K_{ij})$ in the $n$th subcarrier. The conditional utility function is denoted as $U_i(D_{ijn}|K_{ijn})$, that is given by equation (22),

$$U_i(D_{ijn}|K_{ijn}) = \begin{cases} \frac{D_{max,i}(K_{ijn}) - D_{ijn}}{D_{max,i}(K_{ijn}) - \gamma_{D_i}} & D_{ijn} \geq \gamma_{D_i} \\ 1 & D_{ijn} < \gamma_{D_i} \end{cases} \tag{22}$$

$\gamma_{D_i}$ represents the maximum tolerable delay for the $i$th user. $D_{max,i}(K_{ijn}) = \max_n(D_{ijn})$ is the maximum delay for the $i$th user. Additionally, we can infer that $U_i(D_{max,i}|K_{ijn}) = 0$ and $U_i(\gamma_{D_i}|K_{ijn}) = 1$.

When $D_{ijn} < \gamma_{D_i}$, the value of the conditional utility function is equal to 1. Therefore, the total utility function is as follows:

$$\begin{aligned} U_i(D_{ijn}, K_{ijn}) \\ = U_i(D_{ijn}|K_{ijn})U_i(K_{ijn}) \end{aligned} \tag{23}$$

$$U_i(D_{ijn}, K_{ijn}) = \left(1 - \frac{\|X_i(SINR_{ijn}^{UL}) - X_i^R\|}{\max_n \|X_i(SINR_{ijn}^{UL}) - X_i^R\|}\right)\left(\frac{D_{max,i}(K_{ijn}) - D_{ijn}}{D_{max,i}(K_{ijn}) - \gamma_{D_i}}\right) \tag{24}$$

Where $U_i(K_{ijn})$ is the value of utility function for the $i$th user.

### 1.4. OPTIMIZATION PROBLEM

The purpose of the proposed method is to discover the maximum required QoS of users simultaneously in VR systems. The corresponding optimization problem is formulated as equation (25). This problem is the same as the optimal closed form expression for utility function.

$$\max_{U_j, W^{in}, F^{in}} \sum_{j \in B} \sum_{i \in U_j} \sum_{n=1}^{N} U_i(D_{ijn}, K_{ijn}) \tag{25}$$

$$s.t. |U_j| \leq V_j, \forall j \in B \tag{a}$$

$$c_{ij}^{DL} \geq R_{ij}^{min}, \forall j \in B, \forall i \in U_j \tag{b}$$

$$\sum_{i \in U_j} P_{U_i} \leq P_{B_j} \tag{c}$$

$$P_{A_i} \in N_{AP}^j, [P_{A_i}]_{k,l}[P_{A_i}]_{k,l}^* = \frac{1}{N_t} \tag{d}$$

$$G_{A_i} \in N_{STA}^i, [G_{A_i}]_{k,l}[G_{A_i}]_{k,l}^* = \frac{1}{N_r} \tag{e}$$

Where $|U_j|$ is the number of users related to the $j$th AP, and $V_j$ is the total number of users within the coverage of the $j$th AP. Constraint (a) is related to the number of users connected to the $j$th AP. The second constraint denoted that the minimum level of SINR in DL for the $i$th user in contact with the $j$th AP is at-least equal to $R_{ij}^{min}$. As noted in constraint (c), the sum of the power of all users should not exceed the power of the $j$th AP denoted as $P_{B_j}$, where the power $P_{U_i}$ is considered equal for all users. $F^{in} = P_{A_i}^H \hat{P}_{D_{SC(i,n)}}$ represents the precoder matrix, and $W^{in} = G_{A_i} \hat{G}_{D_{SC(i,n)}}$ represents the combiner matrix. Both $G_{A_i}$ and $P_{A_i}$ have unit amplitude. $[P_{A_i}]_{k,l}$ is the element in the $k$th row and $l$th column of the $P_{A_i}$ matrix. $N_{AP}^j$ is the set of codebooks in the $j$th AP, and $[G_{A_i}]_{k,l}$ is the element in the $k$th row and $l$th column of the $G_{A_i}$ matrix. $N_{STA}^i$ is the number of codebooks for the $i$th user.

The maximum value of this utility function is equal to the maximum sum rate of users. The maximization of the sum rate is estimated in the previous section using the SVD algorithm, as explained thoroughly in System Model section. Then by using the values of the rate in DL and UL and queue delay, the utility function is calculated.

## 1.5. SIMULATION RESULTS

In this section, we represent the performance results of the proposed hybrid beamforming algorithm through computer simulations in MATLAB. The environment considered is illustrated in Fig. 2, focusing on an indoor area with two user (U=2) and two AP (B=2). The dimensions of the indoor area is [0:10],[0:17],[0:3] in the X, Y and Z direction, respectively.

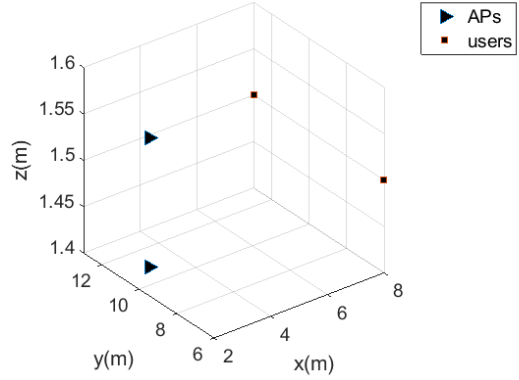

**Fig. 2.** Indoor simulation area.

The values of parameters used in simulations are shown in Table. I. QAM (Quadrature Amplitude Modulation) with 64 subcarriers is considered in OFDM.

TABLE I
PARAMETERS OF SIMULATION

| parameter | value | parameter | value |
|---|---|---|---|
| $f_c$ | 60e9 | $w$ | 3.2 |
| $N_{SC}$ | 64 | $N_r$ | 1 |
| $N_t$ | [2,4,8] | $N_{RF}$ | [1,2] |
| $S_i$ | 512×24 | $A_i$ | 6 |
| $v$ | 5 | $B$ | 2 |
| $U$ | 2 | $P_{B_j}$ | 10e-3 |
| $\mu_j$ | 4e-9 | $\lambda_i$ | 2e-9 |

In order to simulate this network in ns3, we've extract the contents of the PHY layer. At first we have considered six codebooks as follows, for comparing simulation results:
1) "two-Antenna, one-RF chain",
2) "two-Antenna, two-RF chain",
3) "four-Antenna, one-RF chain",
4) "four-Antenna, two-RF chain",
5) "eight-Antenna, one-RF chain",
6) "eight-Antenna, two-RF chain".

Figure. 3 shows the comparison of simulation results in ns3 with simulations in MATLAB in the UL direction in all codebooks for the first user by considering the first AP respectively.

Figure. 4 shows the comparison of simulation results in ns3 with simulations in MATLAB in the UL direction in all codebooks in all codebooks for the first users by considering the second AP respectively.

Similarly Figure. 5 shows the comparison of simulation results in ns3 with simulation in MATLAB in the DL direction in all codebooks for the first users by considering the first AP, respectively.

Similarly Figures. 6 shows the comparison of simulation results in ns3 with simulation in MATLAB in the DL direction in all codebooks for the first user by considering the second AP, respectively.

Consequently, as shown in these figures, the trends of increasing and decreasing UL and DL rate are almost the same as the ns3 result for all considered codebooks for two users and two APs, indicating that the modelling of the channel in DL and UL in our computer simulation is relatively accurate. So we can evaluate the value of our proposed utility function in considered scenarios in the case of number of antennas and RF chains.

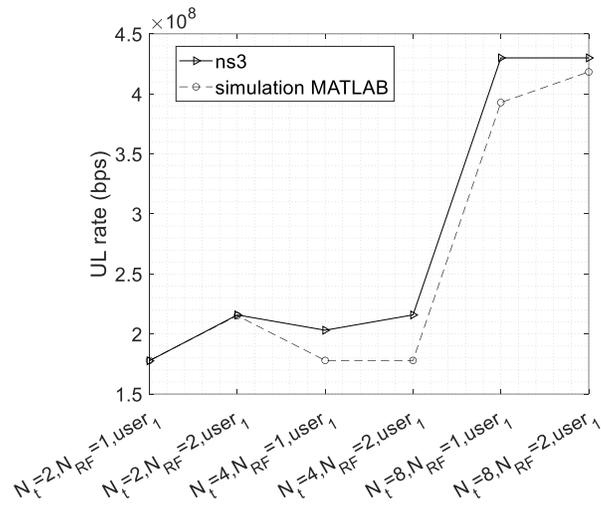

**Fig. 3.** Comparison of UL rate in both ns3 and simulation of MATLAB for the first AP and the first user.

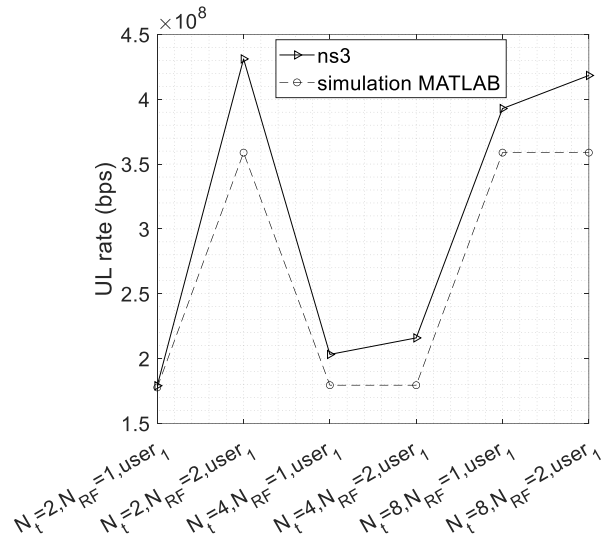

**Fig. 4.** Comparison of UL rate in both ns3 and simulation of MATLAB for the second AP and the first user.

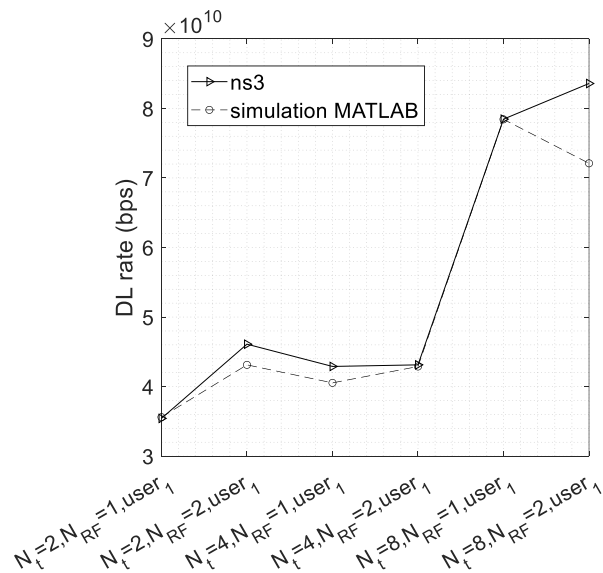

**Fig. 5.** Comparison of DL rate in both ns3 and simulation of MATLAB for the first AP and the first user.

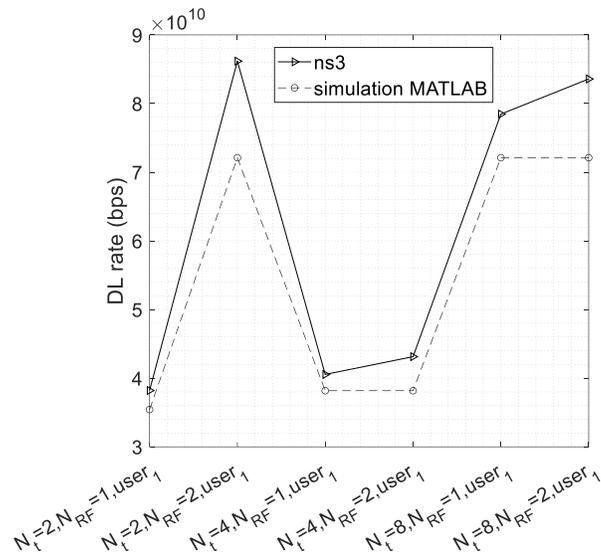

**Fig. 6** Comparison of DL rate in both ns3 and simulation of MATLAB for the second AP and the first user.

Also we have considered lower bound of channel gain in our MATLAB simulation by using the min channel gain in subcarriers. The value of utility function is computed for both of mean channel gain and min channel gain that is shown in Fig.7.

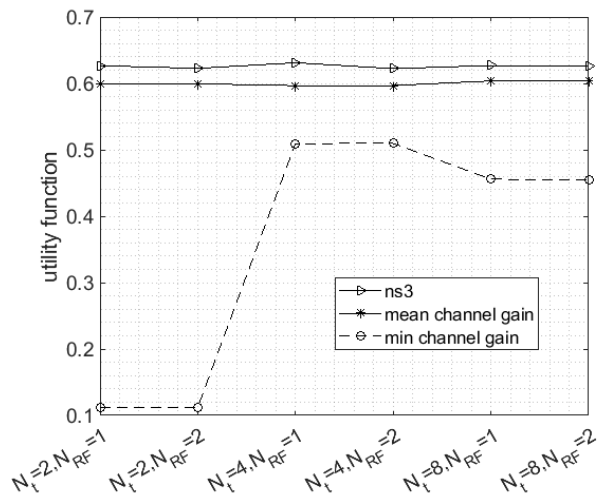

**Fig. 7.** Values of utility function in all scenarios.

In Fig.7, the utility function value for each number of transmit antennas is shown. In the min channel gain scenario, the value of utility function exhibits an increasing trend as the number of antenna is increased. Notably, the values of utility function in the mean channel gain scenario closely align with the simulations in ns3 and the values in all codebooks is approximately equal to 0.6. This alignment is attributed to the consideration of the mean channel gain values in all subcarriers for each users in each Es/N0 value.

As illustrated in Fig 7, ns3 yields higher values of utility function than other scenarios, signifying its superior accuracy in simulating the indoor environment. The average error between ns3 utility function values and the simulation of the first scenario across all codebooks is atleast 4.6%.

In Figs. 8-10, the transmission delay for each Es/N0 (dB) value in both scenarios (mean channel gain and min channel gain), for each codebook, between each user and AP (in cases of varying numbers of $N_t$ and $N_{RF}$) is shown. Almostly the transmission delay exhibits a decreasing pattern with increasing Es/N0 values. This trend is due to the incursion on received power in relation to the power of noise, resulting in an increased rate in DL and a reduction in transmission delay.

*The mean channel gain scenario in MATLAB simulation:*

The transmission delay for some considered codebooks in the mean channel gain scenario is illustrated in Figures.8 through 10.

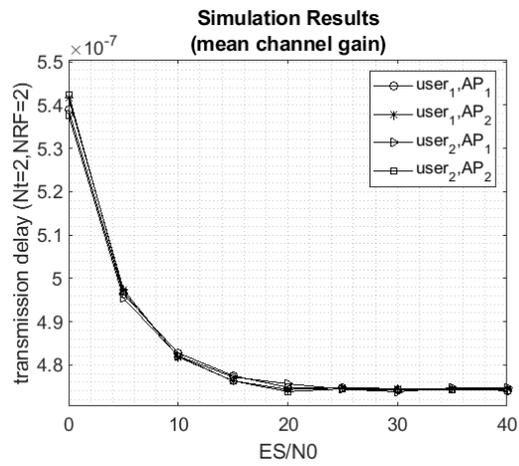

**Fig. 8.** Transmission delay in mean channel gain scenario for $N_t = 2, N_{RF} = 2$.

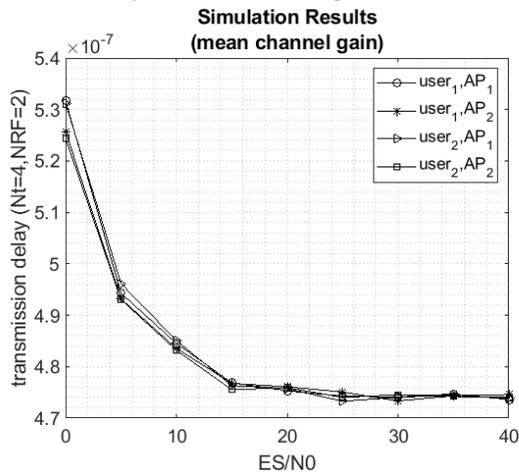

**Fig. 9.** Transmission delay in mean channel gain scenario for $N_t = 4, N_{RF} = 2$.

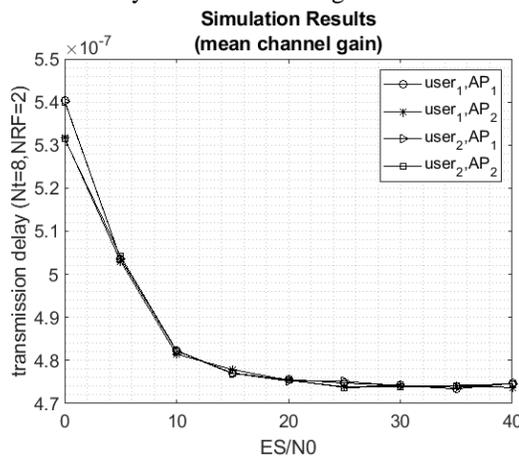

**Fig. 10** Transmission delay in mean channel gain scenario for $N_t = 8, N_{RF} = 2$.

*MATLAB simulations of min channel gain scenario:*

The transmission delay for some considered codebooks in the min channel gain scenario is illustrated in Figures.11 through 13.

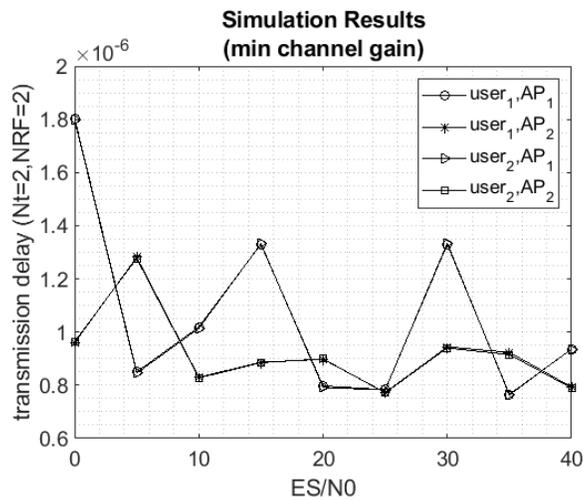

**Fig. 11.** Transmission delay in min channel gain scenario for $N_t = 2, N_{RF} = 2$.

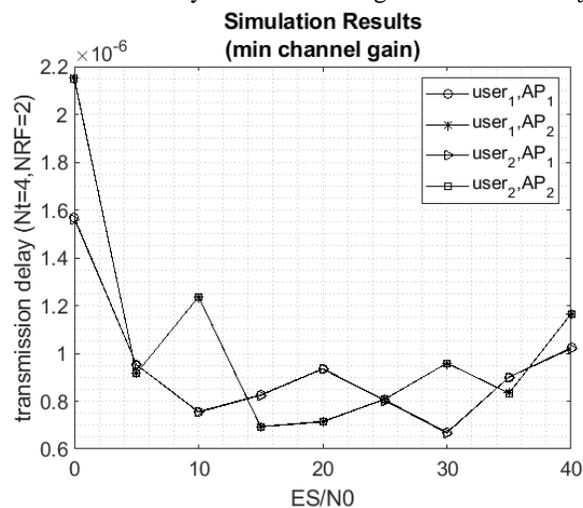

**Fig. 12.** Transmission delay in min channel gain scenario for $N_t = 4, N_{RF} = 2$.

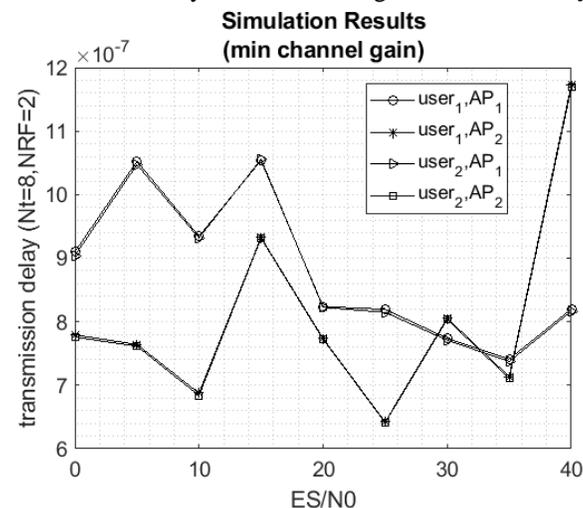

**Fig. 13.** Transmission delay in min channel gain scenario for $N_t = 8, N_{RF} = 2$.

After simulating the transmission delay in various Es/N0 values across scenarios, the min and mode values of transmission delay at different Es/N0 values are compared with each other and with the ns3 results. Mode statistics pertains to the most frequently occurring value in a vector. The minimum and mode values of transmission delay for all considered codebooks within the range of Es/N0 values in all scenarios are shown in Figs. 14-17.

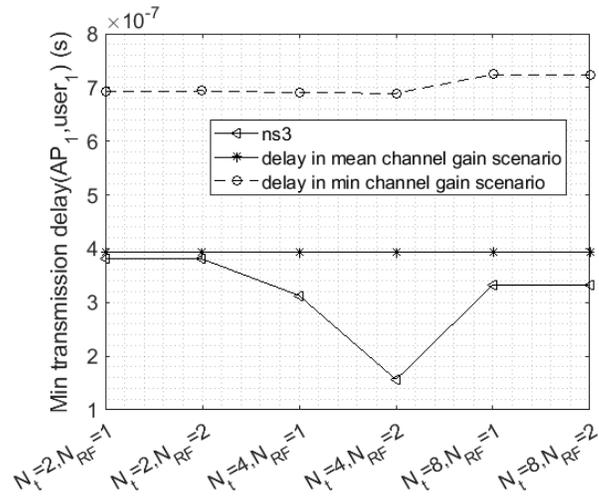

**Fig. 14.** Min transmission delay based on all Es/N0 values in scenarios and ns3 for the first AP and the first user.

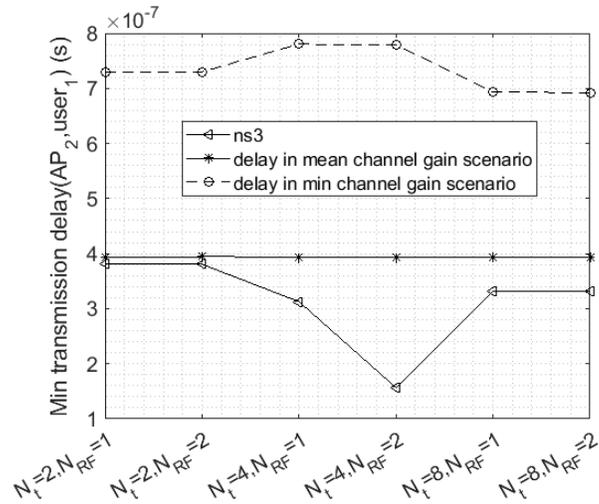

**Fig. 15.** Min transmission delay based on all Es/N0 values in scenarios and ns3 for the second AP and the first user.

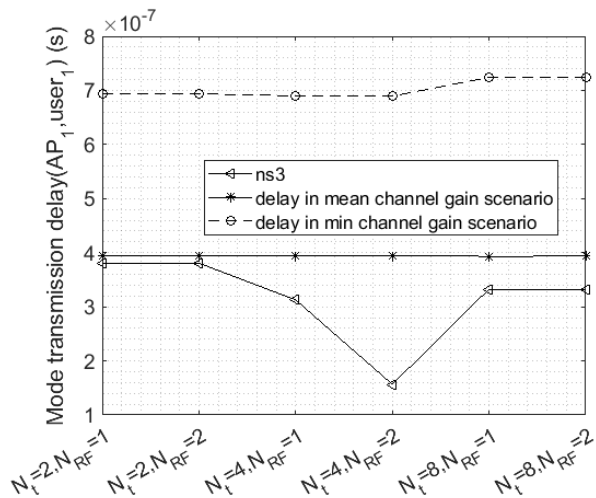

**Fig. 16.** Mode of transmission delay based on all Es/N0 values in scenarios and ns3 for the first AP and the first user.

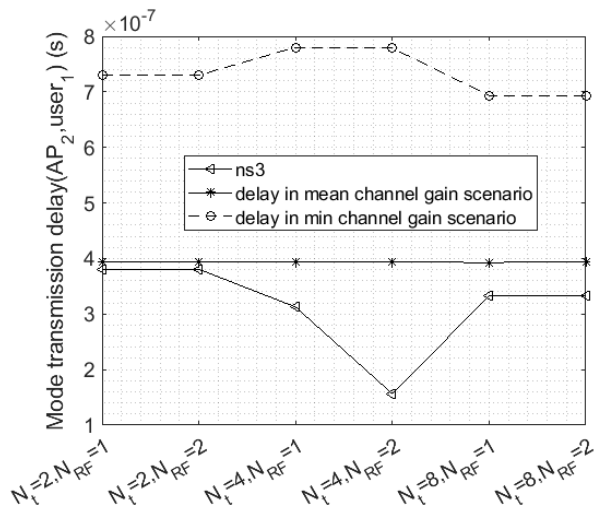

**Fig. 17.** Mode of transmission delay based on all Es/N0 values in scenarios and ns3 for the second AP and the first user.

As shown in Fig. 14 and Fig.15, ns3 consistently exhibits the lowest transmission delay across all considered codebooks. This outcome is attributed to the higher DL and UL rates obtained in ns3, surpassing the simulated rates in the mean channel gain scenario using MATLAB. The minimum transmission delay in the min channel gain scenario has higher values. In the min channel gain scenario, the consideration of the minimum value of the channel at subcarriers in DL for calculating the DL rates contributes to higher transmission delay.

Figs.16-17 have shown the mode value of transmission delay that are selected from all transmission delay in all values of Es/N0. The mode value of transmission delay in the min channel gain scenario is higher than others, indicating the presence of outliers in transmission delay for this scenario. AS shown in these figures, the mode of transmission delay is almost the same as the minimum value shown in Figs.14-15. Interestingly, the mode of transmission delay in the mean channel gain scenario exhibits approximately the same values as the minimum transmission values based on Figs.14-15 and closely approaches ns3 results in Figs.16-17.

### 1.6. CONCLUSION

In this paper, we have focused on hybrid beamforming design within mmWave multi-carrier systems in DL to enhance the received power of users while minimizing hardware complexity. Additionally, we proposed an optimal closed form expression of a multi-attribute utility function to assess the QoS of users, taking into account the specific limitations of VR systems. This multi attribute utility function includes transmission delay, processing delay and queue delay. We investigate the transmission delay in relation to increasing Es/N0 and the utility function for mean channel gain and min channel gain scenarios in six different codebooks with various number of transmit antenna and RF chains in DL. Furthermore, we have discussed a lower bound consideration on channel gain that can lead to a reduction in DL rate and ultimately increasing transmission delay. Consequently, the utility function values have decreased by considering the lower bound of channel gain.

The results reveal a consistent trend in the utility function value for each codebook and on average, the accuracy of utility function in ns3 is about 4.6% in comparison by MATLAB simulations across all considered codebooks.